\documentclass[final,times,twocolumn]{elsarticle}
\usepackage{graphicx,amssymb,amsmath,hyperref}
\usepackage{multirow,dcolumn,bm,latexsym,soul}
\newcolumntype{K}[1]{>{\centering\arraybackslash}p{#1}}

\journal{Physics of the Dark Universe}
\begin{document}
\newcommand{\be}{\begin{equation}}
\newcommand{\ee}{\end{equation}}
\newcommand{\bq}{\begin{eqnarray}}
\newcommand{\eq}{\end{eqnarray}}
\newcommand{\bw}{\begin{widetext}}
\newcommand{\ew}{\end{widetext}}
\newcommand{\bsq}{\begin{subequations}}
\newcommand{\esq}{\end{subequations}}
\newcommand{\bc}{\begin{center}}
\newcommand{\ec}{\end{center}}
\begin{frontmatter}

\title{Observational constraints on nonlinear matter extensions of general relativity: Separable trace power models}
\author[inst1]{E.-A. Kolonia}\ead{eleannakolonia@gmail.com}
\author[inst2,inst3]{C. J. A. P. Martins\corref{cor1}}\ead{Carlos.Martins@astro.up.pt}
\address[inst1]{Department of Physics, University of Patras, 26504 Patras, Greece}
\address[inst2]{Centro de Astrof\'{\i}sica da Universidade do Porto, Rua das Estrelas, 4150-762 Porto, Portugal}
\address[inst3]{Instituto de Astrof\'{\i}sica e Ci\^encias do Espa\c co, CAUP, Rua das Estrelas, 4150-762 Porto, Portugal}
\cortext[cor1]{Corresponding author}

\begin{abstract}
{The search for the physical mechanism underlying the observational evidence for the acceleration of the recent universe is a compelling goal of modern fundamental cosmology. Here we quantitatively study a class of homogeneous and isotropic cosmological models in which the matter side of Einstein's equations includes, in addition to the canonical term, a term proportional to the trace of the energy-momentum tensor, $T=\rho-3p$, and constrain these models using low redshift background cosmology data. One may think of these models as extensions of general relativity with a nonlinear matter Lagrangian, and they can be studied either as phenomenological extensions of the standard $\Lambda$CDM model, containing both matter and a cosmological constant, or as direct alternatives to it, where there is no cosmological constant but the additional terms would have to be responsible for accelerating the universe. Overall, our main finding is that parametric extensions of $\Lambda$CDM are tightly constrained, with additional model parameters being constrained to their canonical behaviours to within one standard deviation, while alternative models in this class (which do not have a $\Lambda$CDM limit) are ruled out. This provides some insight on the level of robustness of the $\Lambda$CDM model and on the parameter space still available for phenomenological alternatives and extensions.}
\end{abstract}
\end{frontmatter}

\section{Introduction}\label{introd}

Since the discovery of the observational evidence for the acceleration of the universe, there has been a systematic effort to identify its underlying physical mechanism \cite{Copeland,Frieman}. Among the many classes of models that have been put forward, one that has been the subject of recent interest is a class of Friedmann-Lema\^{\i}tre-Robertson-Walker (FLRW) models in which the matter side of Einstein's equations includes, in addition to the canonical term, a further term proportional to some function of the energy-momentum tensor ($T^2=T_{\alpha\beta}T^{\alpha\beta}=\rho^2+3p^2$), or of its trace ($T=\rho-3p$). Qualitatively, one may think of these models as extensions of general relativity with a nonlinear matter Lagrangian. This makes them phenomenologically interesting because they are somewhat different from the usual dynamical dark energy or modified gravity models, in the sense that in the former class of models one adds further dynamical degrees of freedom to the Lagrangian (often in the form of scalar fields), while in the latter the gravitational part of the Lagrangian is changed. 

One recent example of the first class of models is provided by the work of \cite{Roshan}, who studied a model of the so-called energy-momentum-squared gravity, where the matter part of Einstein's equations is modified by the addition of a term proportional to $T^2$. Subsequent works \cite{Board,Akarsu} have extended this to the more generic form $(T^2)^n$, dubbed energy-momentum-powered gravity. The work of \cite{Faria}, and more recently \cite{Eleanna}, provided low redshift constraints on these models, from the combination of two different datasets.

In practical phenomenological terms, we may think of these models as extensions to the canonical $\Lambda$CDM: the model still has a standard cosmological constant (except if one stipulates that it should vanish) but the nonlinear matter Lagrangian leads to additional terms in Einstein's equations, and cosmological observations can therefore constrain the corresponding additional model parameters which appear in these terms. Typically there are two such additional parameters: the power $n$ of the nonlinear part of the Lagrangian, and a further parameter (to be defined below) quantifying the contribution of this term to the energy budget of the universe. The particular choice of $n=0$ does correspond to $\Lambda$CDM, so effectively these models are always parametric extensions thereof, and the two additional model parameters are therefore tightly constrained. Specifically, the analyses of \cite{Faria,Eleanna} found that these models do not solve the cosmological constant problem {\it per se}, but they can phenomenologically lead to a recent accelerating universe at the cost of having preferred values of the cosmological parameters, such as the matter density or its equation of state parameter, that are somewhat different from the standard $\Lambda$CDM ones.

One may also ask the more generic question of whether a suitably chosen nonlinear Lagrangian can reproduce the recent (low redshift) acceleration of the universe in a model which at low redshift only contains matter (plus a subdominant amount of radiation) but no true cosmological constant. In principle such a scenario is conceivable and has been qualitatively discussed in the original work \cite{Roshan}. It is also somewhat closer in spirit to the usual modified gravity models with the caveat that, as previously mentioned, in the latter models the modification occurs in the gravitational part of the Lagrangian and not in the matter part.

One way to address this question is to consider a second class of models, corresponding to the case where the new terms depend on the trace of the energy-momentum tensor $T=\rho-3p$, or indeed some power thereof. Such models have been recently considered \cite{Velten,Godani}, and some quantitative low-redshift constraints on a comparatively simple model within this class have recently been presented in \cite{Eleanna}. Here we build upon this work, take these phenomenological models at face value, and present a more extensive analysis of this class of models, comparing them to the same two previously mentioned datasets. In particular, we consider both the general scenario with a cosmological constant (in which case the model is an extension of $\Lambda$CDM) and the scenario without a cosmological constant (in which case we can check whether such models can accelerate at all), as well as some additional specific examples of models in this class. In our analysis the Hubble constant was analytically marginalized as discussed in \cite{Anagnostopoulos}.

The first dataset is the Pantheon Type Ia supernova compilation \cite{Riess}. This is a 1048 supernova dataset, containing measurements in the range $0.01<z<2.3$, further compressed into 6 correlated measurements of $E^{-1}(z)$ (where $E(z)=H(z)/H_0$ is the dimensionless Hubble parameter) in the redshift range $0.07<z<1.5$. This compression is specifically introduced and discussed in \cite{Riess}, which shows that it provides an effectively identical characterization of the expansion history and dark energy as the full supernova sample, thus making it an efficient compression of the raw data.\footnote{The Pantheon repository \url{https://github.com/dscolnic/Pantheon} contains a subsequent revision of the heliocentric and cosmological redshifts of the two datasets. Comparing the latter, expressed as the modulus of the relative difference, one finds that the maximum difference is $1.01\%$, the mean difference is $0.07\%$ and the median difference is $0.08\%$. Clearly these differences are too small to impact our results.}

The second dataset is a compilation of 38 Hubble parameter measurements of Faroon \textit{et al.} \cite{Farooq}, which includes both baryon acoustic oscillation and cosmic chronometer data. Regarding the latter, we note that it has been argued that possible systematics issues of the method may not yet be well understood and under control \cite{Concas,Vazdekis}. A recent detailed discussion can be found in \cite{Moresco}. We note that among the cosmic chronometers measurements listed in Table 1 of \cite{Moresco} there are two that are not part of the Farooq \textit{et al.} \cite{Rats,Borghi}). However, these two are among the cosmic chronometer measurements with larger error bars, and we have checked that including them in our analysis would not significantly change our constraints. This is due to the fact that the constraining power of the cosmic chronometers is weaker than that of the baryon acoustic oscillations and supernova measurements.

The plan of this paper is as follows. We start in Sect. \ref{mods} with a brief overview of our recent results on $(T^2)^n$ and $\sqrt{T}$ models, also due to be published in a conference's proceedings \cite{Eleanna}, with the dual aims of introducing some of the necessary notation and of providing a comparison point for the results we discuss subsequently. In Sect. \ref{fullmod} we go beyond these earlier results, first by introducing the general trace power ($T^n$) models and then by discussing some particular solutions thereof. Section \ref{partconst} then presents constraints on the particular solutions of the model for which the continuity equation is analytically integrable, while Sect. \ref{fullconst} presents constraints on the general model, under various assumptions. In passing we note that specific cases of $(T^2)^n$ and $T^n$ models also have a superficial resemblance to a simpler toy model known as the Cardassian model \cite{Cardassian1}, and we briefly comment on this in Sect. \ref{Card}. Finally, Sect. \ref{concl} contains a summary and discussion of our results.

\section{Overview of $(T^2)^n$ and $\sqrt{T}$ models}\label{mods}

The general action for energy-momentum-powered models is \cite{Roshan,Board}
\be
S=\frac{1}{2\kappa}\int\left[R+\eta (T^2)^n-2\Lambda\right]d^4x + S_{matter}\,,
\ee
where $\kappa=8\pi G$, $\Lambda$ is the cosmological constant, and $\eta$ is a constant quantifying the contribution of the $T^2$-dependent term. In a flat Friedmann-Lemaitre-Robertson-Walker universe and assuming a perfect fluid we have $T^2=\rho^2+3p^2$ and the Friedmann, Raychaudhuri and continuity equations can be written as
\be
3\left(\frac{\dot a}{a}\right)^2=\Lambda+\kappa\rho+\eta(\rho^2+3p^2)^{n-1}\left[\left(n-\frac{1}{2}\right)(\rho^2+3p^2)+4np\rho\right]
\ee
\be
6\frac{\ddot a}{a}=2\Lambda-\kappa(\rho+3p)-\eta(\rho^2+3p^2)^{n-1}\left[(n+1)(\rho^2+3p^2)+4np\rho\right]
\ee
\be
{\dot\rho}=-3\frac{\dot a}{a}(\rho+p)F(n,\eta,\rho,p),
\ee
where for convenience we have defined
\be
F(n,\eta,\rho,p)=\frac{\kappa\rho+n\eta\rho(\rho+3p)(\rho^2+3p^2)^{n-1}}{\kappa\rho+n\eta(\rho^2+3p^2)^{n-1}\left[\left(2n-1\right)(\rho^2+3p^2)+8np\rho\right]}.
\ee
As usual, only two of these equations are independent; for our purposes in the present work, the most convenient choice is to use the Friedmann and continuity equations. Note that for either $\eta=0$ or $n=0$ one recovers $\Lambda$CDM,

In general these equations need to be solved numerically. However, there are three particular cases for which analytic solutions can be found (at least approximate, low redshift solutions), corresponding to the values $n=1$, $n=1/2$ and $n=0$. These have been studied in a general mathematical context \cite{Roshan,Early,Board}, and have also been observationally constrained in \cite{Faria}.

Generically we can treat $n$ as a free parameter, to be constrained by observations. We define a dimensionless cosmological density $r$, via $\rho = r \rho_0$, where $\rho_0$ is the present day density, as well as a generic parameter
\be
Q=\frac{\eta}{\kappa}\rho_0^{2n-1}\,.
\ee
We can also consider a further generalization: instead of stipulating a universe with a matter fluid, we can assume that this fluid has a constant equation of state parameter, $w=p/\rho=const.$ (with the matter case corresponding to $w=0$). Nevertheless, one should bear in mid that there are strong observational constraints in the matter equation of state, as discussed for example in \cite{Tutusaus}. With these assumptions the continuity equation can be written
\be
\frac{dr}{dz}=\frac{3r}{1+z}(1+w)\times\frac{1+nQf_1(n,w)r^{2n-1}}{1+2nQf_2(n,w)r^{2n-1}}\,,
\ee
where for convenience we defined
\be
f_1(n,w)=(1+3w)(1+3w^2)^{n-1}\,, \label{deff1}
\ee
\be
f_2(n,w)=(1+3w^2)^{n-1}\left[\left(n-\frac{1}{2}\right)(1+3w^2)+4nw\right]\,. \label{deff2}
\ee
On the other hand the Friedmann equation can be written
\be
E^2(z)=\Omega_\Lambda+\Omega_Mr+f_2(n,w)Q\Omega_Mr^{2n}\,,
\ee
together with the consistency relation $\Omega_\Lambda=1-\Omega_M[1+f_2Q]$. Two alternative ways of writing it are
\be
E^2(z)=\Omega_\Lambda+\Omega_Mr+(1-\Omega_M-\Omega_\Lambda)r^{2n}\,
\ee
\be
E^2(z)=1+\Omega_M(r-1)+f_2(n,w)Q\Omega_M(r^{2n}-1)\,,
\ee
where the first is generic while the second holds for $\Omega_\Lambda\neq0$. Conversely, if $\Omega_\Lambda=0$ the continuity equation can also be written in a way that eliminates $Q$,
\be\label{fullnoq}
\frac{dr}{dz}=\frac{3r}{1+z}(1+w)\times\frac{\Omega_Mf_2+n(1-\Omega_M)f_1r^{2n-1}}{f_2[\Omega_M+2n(1-\Omega_M)r^{2n-1}]}\,.
\ee

\begin{table*}
\begin{center}
\caption{One sigma posterior likelihoods on the matter density $\Omega_M$, the power $n$ and the constant equation of state parameter $w$ (when applicable) for various flat energy-momentum-powered models, with or without a cosmological constant. The last column lists the reduced chi-square for each best-fit model. The constraints come from the combination of the Pantheon supernova data and Hubble parameter measurements. These have also been reported in a recent conference proceedings \cite{Eleanna}.}
\label{table1}
\begin{tabular}{| c | c | c | c | c |}
\hline
Model assumptions & $\Omega_M$ & $n$  & $w$ & $\chi^2_\nu$ \\
\hline
$\Omega_\Lambda=0$, $w=0$ & $0.39\pm0.08$ & $0.04\pm0.04$ & N/A & 0.64\\
\hline
$\Omega_\Lambda\neq0$, $w=0$ & $0.29^{+0.05}_{-0.03}$ & Unconstrained & N/A & 0.64 \\
\hline
$\Omega_\Lambda=0$, $w=const.$  & $0.28^{+0.12}_{-0.10}$ & $-0.08^{+0.06}_{-0.02}$& $-0.11^{+0.07}_{-0.04}$ & 0.62 \\
\hline
\end{tabular}
\end{center}
\end{table*}

Table \ref{table1} summarizes the observational constraints on these models from the previously described datasets, for various assumptions on the presence or absence of a cosmological constant and on the allowed equation of state parameter. These various cases are discussed in more detail in \cite{Eleanna}. The salient points are the significant degeneracies between model parameters and the fact that that these models overfit the data, with values of the reduced chi-square (i.e., the chi-square per degree of freedom) around $\chi^2_\nu\sim0.6$. This is far smaller than the value of the reduced chi-square obtained by comparing the standard CPL parametrization to the same two datasets, which is $\chi^2_\nu\sim0.9$ as reported in \cite{Fernandes}. This shows that these models compare disfavourably to the more standard parametrization. The best-fit values of the model parameters are about one standard deviation away from the canonical values $n=0$ and $\Omega_M\sim0.3$, but at the two sigma level the results are consistent with $\Lambda$CDM.

It is interesting to contrast this class of models with the one where the new terms depend on the trace of the energy-momentum tensor $T=\rho-3p$, rather than $T^2=\rho^2+3p^2$. One convenient example of the latter class is the modified gravity model recently discussed in \cite{Godani}, and also previously considered in \cite{Velten}. This is actually one case of a larger set of models, to be discussed in the next section. The model has the action
\be
S=\frac{1}{2\kappa}\int\left[R+\xi \sqrt{T}-2\Lambda\right]d^4x + S_{matter}\,,
\ee
In a flat FLRW universe the corresponding Friedmann and Raychaudhuri equations are
\be
3\left(\frac{\dot a}{a}\right)^2=\Lambda+\kappa\rho+\xi\frac{(\rho-p)}{\sqrt{\rho-3p}}
\ee
\be
6\frac{\ddot a}{a}=2\Lambda-\kappa(\rho+3p)+\frac{\xi}{2}\frac{(\rho-7p)}{\sqrt{\rho-3p}}\,.
\ee

As a simple but interesting comparison between the two classes of models, in the $p=0$ case this model leads to a Friedmann equation
\be
3H^2=\Lambda+\kappa\rho+\xi\sqrt{\rho}\,,
\ee
while in the energy-momentum-powered model, choosing $n=1/4$, one has
\be
3H^2=\Lambda+\kappa\rho-\frac{\eta}{4}\sqrt{\rho}\,.
\ee
Thus the Friedmann equations in the two models coincide (if one identifies $\xi=-\eta/4$), but the corresponding continuity equations differ in the two cases, as discussed presently.

In general, and as in the previous case, we will assume a constant equation of state parameter ($p=w\rho$), use $\rho=r\rho_0$ and additionally define
\be\label{defzeta}
\zeta=\frac{\xi}{2\kappa\sqrt{\rho_0}}
\ee
With these definitions we can rewrite the Friedmann equation as follows
\be
E^2(z)=\Omega_\Lambda+\Omega_Mr+2\zeta\frac{(1-w)}{\sqrt{1-3w}}\Omega_M\sqrt{r}\,.
\ee
In principle we now have 3 free parameters, since the $E(0)=1$ condition requires that  $\Omega_\Lambda=1-\Omega_M[1+2\zeta(1-w)/\sqrt{1-3w}]$. In the general case this can be written
\be
E^2(z)=\Omega_\Lambda+\Omega_Mr+(1-\Omega_M-\Omega_\Lambda)\sqrt{r}\,,
\ee
while if $\Omega_\Lambda\neq0$ we can write
\be
E^2(z)=1+\Omega_M(r-1)+2\zeta\frac{(1-w)}{\sqrt{1-3w}}\Omega_M(\sqrt{r}-1)\,,
\ee
but note that in general the parameters $(\zeta,w)$ still affect the continuity equation, which can be written
\be
\frac{dr}{dz}=\frac{3r}{1+z}(1+w)\times\frac{\sqrt{1-3w}+\zeta/\sqrt{r}}{\sqrt{1-3w}+(1-w)\zeta/\sqrt{r}}\,.
\ee
As expected, the usual behaviour, $r\propto (1+z)^3$, is recovered for $\zeta=0$. Less trivially, this also occurs for the matter case ($w=0$) for any value of the parameter $\zeta$---which is a significant difference with respect to the energy-momentum-powered case. To illustrate the role of this parameter we can solve the continuity equation in the $\zeta\longrightarrow0$ limit, finding
\be
r(z)=\left[\left(1+\frac{w\zeta}{\sqrt{1-3w}}\right)(1+z)^{3(1+w)/2}-\frac{w\zeta}{\sqrt{1-3w}}\right]^2\,,
\ee
which again has the appropriate limits.

Table \ref{table2} shows the analogous constraints for this model, also recently discussed in a conference proceedings \cite{Eleanna}, for the same three scenarios already discussed for the energy-momentum powered case and also for the same datasets. For $\Omega_\Lambda=0$ and $w=0$ (first row of the table) there is only one independent parameter, since the matter density and the coupling are related via $(1+2\zeta)\Omega_M=1$. The very large value of the reduced chi-square shows that this does not fit the data.

\begin{table*}
\begin{center}
\caption{One sigma posterior likelihoods on the matter density $\Omega_M$, the coupling $\zeta$ and the constant equation of state parameter $w$ (when applicable) for various flat $\sqrt{T}$ models, with or without a cosmological constant. The last column lists the reduced chi-square for each best-fit model. The constraints come from the combination of the Pantheon supernova data and Hubble parameter measurements. These have also been reported in a recent conference proceedings \cite{Eleanna}. Note that for the first model $\zeta$ is not independent from $\Omega_M$.}
\label{table2}
\begin{tabular}{| c | c | c | c | c |}
\hline
Model assumptions & $\Omega_M$ & $\zeta$ & $w$ & $\chi^2_\nu$ \\
\hline
$\Omega_\Lambda=0$, $w=0$ & $0.15\pm0.02$ & ($2.78\pm1.72$) & N/A & 1.80\\
\hline
$\Omega_\Lambda\neq0$, $w=0$ & $0.25^{+0.03}_{-0.02}$ & $0.23^{+0.22}_{-0.18}$ & N/A & 0.63 \\
\hline
$\Omega_\Lambda=0$, $w=const.$ & $0.24^{+0.08}_{-0.07}$ & Unconstrained & $-0.08^{+0.04}_{-0.05}$ & 1.80 \\
\hline
\end{tabular}
\end{center}
\end{table*}

\begin{figure*}
\begin{center}
\includegraphics[width=\columnwidth]{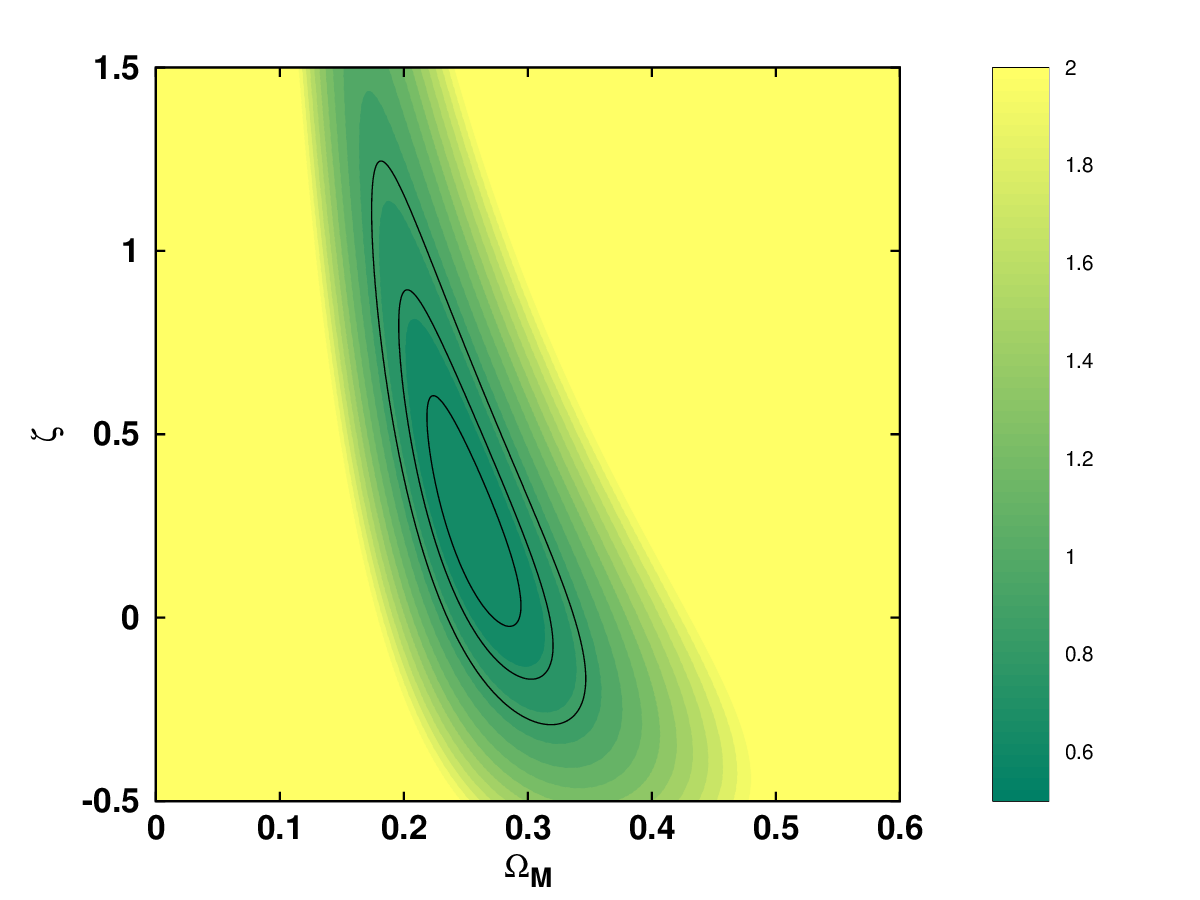}
\includegraphics[width=\columnwidth]{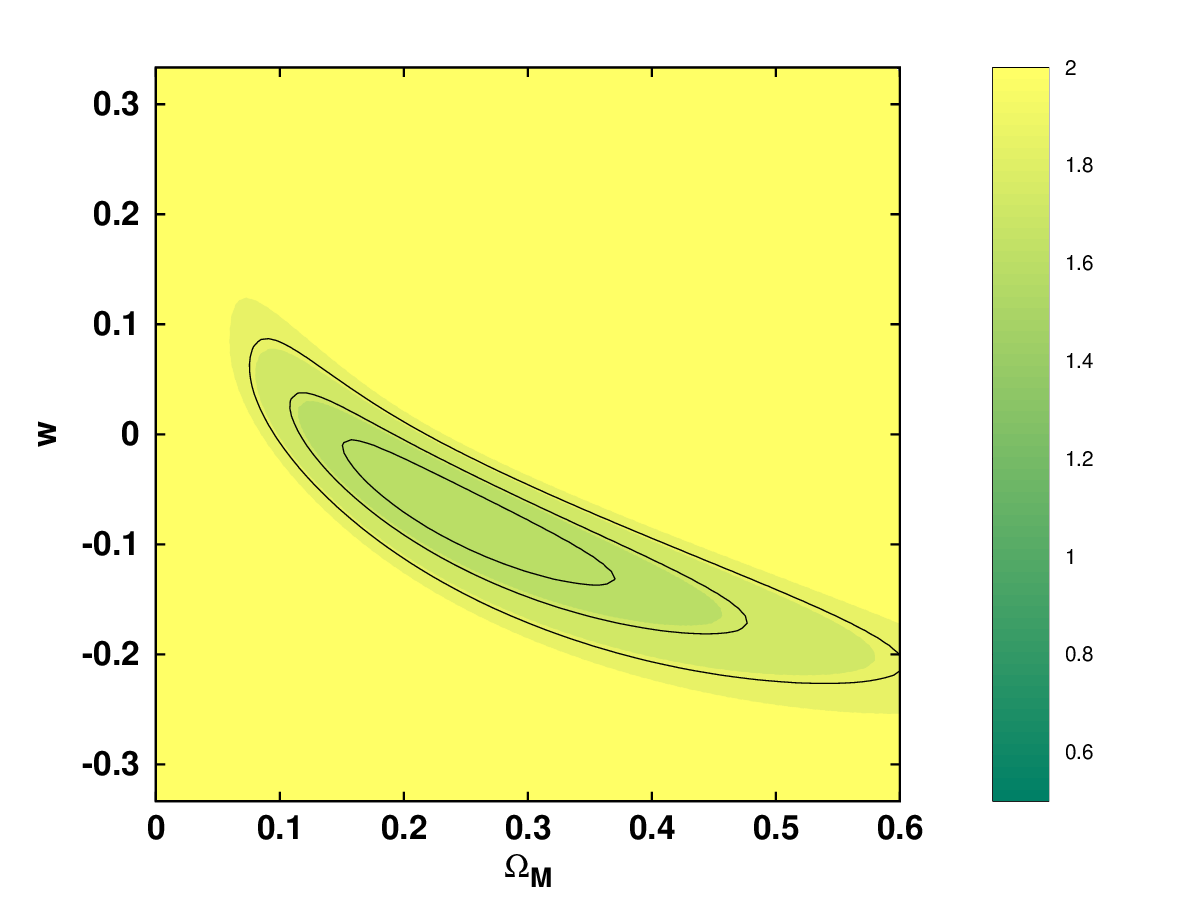}
\end{center}
\caption{Constraints on flat $\sqrt{T}$ models. The left panel shows constraints for $\Omega_\Lambda\neq0$ and $w=0$, and the right panel shows constraints for $\Omega_\Lambda=0$ and $w=const$. The black solid curves show the one, two, and three sigma confidence levels, and the color maps depict the reduced chi-square. Similar constraints are shown in Figure 3 of \cite{Eleanna}.}
\label{figure01}
\end{figure*}

The left panel of Fig. \ref{figure01} and the middle row of the table summarize the constraints for $\Omega_\Lambda\neq0$ and $w=0$. Here there are two independent parameters, and the model is effectively a one parameter extension of $\Lambda$CDM. As in the energy-momentum-powered case, the model overfits the data, but there is no statistically significant preference for a non-zero coupling parameter $\zeta$, Finally, the right panel of Fig. \ref{figure01} and the bottom row of Table \ref{table2} show the constraints for $\Omega_\Lambda=0$ and $w\neq0$, in which case there are three independent parameters. Here the outcome is the same as in the first case: without a cosmological constant this model severely underfits the data, and therefore it is not viable as an alternative to $\Lambda$CDM.

\section{General $T^n$ models}\label{fullmod}

A class of modified gravity models now dubbed $f(R)$ gravity, where $R$ denotes the scalar curvature, was first considered in \cite{Buchdahl}, but these models are now subject to tight cosmological constraints \cite{Amendola1,Amendola2,Clifton}. A phenomenologically  broader (if physically less well motivated) class is that of the so-called $f(R,T)$ models \cite{Harko}, where $T$ is the trace of the stress energy tensor. A particular subclass of the latter models has separable function, $f(R,T)=f_1(R)+f_2(T)$, and in what follows we further set $f_1(R)=R$. These models have been the subject of several mathematical studies but so far they have not been put through a detailed comparison with cosmological observations, with the exception of the recent qualitative analysis of \cite{Velten}.

One can show \cite{Velten} that for models with $f(R,T)=R+f_2(T)$, and again further assuming a constant equation of state parameter $p=w\rho$, the Friedmann and Raychaudhuri equations can be written
\be
3\left(\frac{\dot a}{a}\right)^2=\Lambda+\kappa\rho+\frac{1}{2}f_2+(1+w)\rho f'_2
\ee
\be
6\frac{\ddot a}{a}=2\Lambda-\kappa(1+3w)\rho+f_2-(1+w)\rho f'_2\,,
\ee
where the prime denotes $f'_2=df_2(T)/dT$. From these one obtains the continuity equation
\be
{\dot \rho}+3H(1+w)\frac{\kappa+f'_2}{\kappa+(1+w)f'_2}\rho=-\frac{{\dot f_2}/2+(1+w){\dot f'_2}\rho}{\kappa+(1+w)f'_2}\,.
\ee

In the present work we take $f_2(T)=T^n$, and therefore our action is
\be
S=\frac{1}{2\kappa}\int\left[R+\theta T^n-2\Lambda\right]d^4x + S_{matter}\,,
\ee
where again $T=\rho-3p=(1-3w)\rho$. The model studied in the previous section corresponds to $n=1/2$ and $\theta=\xi$, while here we treat the exponent $n$ and the coupling $\theta$ as free phenomenological parameters, to be constrained by data. In this case the Friedmann and Raychaudhuri equations become
\be
3\left(\frac{\dot a}{a}\right)^2=\Lambda+\kappa\rho+\frac{\theta}{2}[1+2n+(2n-3)w](1-3w)^{n-1}\rho^n
\ee
\be
6\frac{\ddot a}{a}=2\Lambda-\kappa(1+3w)\rho+\theta[1-n-(n+3)w](1-3w)^{n-1}\rho^n\,,
\ee
while the continuity equation becomes
\be
{\dot \rho}=-3H(1+w)\frac{\kappa+\theta n(1-3w)^{n-1}\rho^{n-1}}{\kappa+\theta n[1/2+n+(n-3/2)w](1-3w)^{n-1}\rho^{n-1}}\,,
\ee
and again we notice that with the choices $\theta=0$ or $n=0$ we recover the $\Lambda$CDM case, as expected.

Further defining
\be\label{defbeta}
\beta=\frac{\theta}{2\kappa}\rho_0^{n-1}\,,
\ee
which generalizes Eq. (\ref{defzeta}), we can rewrite the Friedmann equation as
\be
E^2(z)=\Omega_\Lambda+\Omega_Mr+\beta[1+2n+(2n-3)w](1-3w)^{n-1}\Omega_Mr^n\,,
\ee
where again the model's free parameters are related by the consistency condition $E(0)=1$, which one can use to simplify the Friedmann equation to
\be
E^2(z)=\Omega_\Lambda+\Omega_Mr+(1-\Omega_M-\Omega_\Lambda)r^{n}\,.
\ee
In the particular case of $\Omega_\Lambda\neq0$ we can also write it in the convenient form
\be
E^2(z)=1+\Omega_M(r-1)+\beta[1+2n+(2n-3)w](1-3w)^{n-1}\Omega_M(r^{n}-1)\,.
\ee
The continuity equation, which we use in our numerical analysis together with the Friedmann equation, can be written
\be
\frac{dr}{dz}=\frac{3r}{1+z}(1+w) G(\beta,n,w,r)\,,
\ee
where for convenience we have defined
\be
G(\beta,n,w,r)=\frac{1+2\beta n(1-3w)^{n-1}r^{n-1}}{1+\beta n[1+2n+(2n-3)w](1-3w)^{n-1}r^{n-1}}\,.
\ee
It is straightforward to check that these reduce to those of the previous section when $n=1/2$.

It is interesting to note that the standard behaviour of the continuity equation, $r\propto (1+z)^{3(1+w)}$, is recovered provided one has $\beta=0$ (which corresponds to the standard model if a cosmological constant is allowed), $w=1/3$, $n=0$, or for the particular choice \cite{Alvarenga,Velten}
\be
n=\frac{1+3w}{2(1+w)}\,;\label{nwrelation}
\ee
this expression also highlights the particular status of $n=1/2$ for $w=0$. Indeed, for small $w$ \cite{Tutusaus} one can Taylor-expand to obtain $n=w+1/2$. In what follows we first discuss constraints for these special cases, and then consider the general case.

\begin{table*}
\begin{center}
\caption{One sigma posterior likelihoods on the matter density $\Omega_M$, the constant equation of state parameter $w$ and the coupling $\beta$ (when applicable) for various particular cases of the $T^n$ model, discussed in Sect. \ref{partconst}. The last column lists the reduced chi-square for each best-fit model. The constraints come from the combination of the Pantheon supernova data and Hubble parameter measurements.}
\label{table3}
\begin{tabular}{| c | c | c | c | c |}
\hline
Case  & $\Omega_M$ & $w$ & $\beta$ & $\chi^2_\nu$ \\
\hline
Case 1 ($w=1/3$) & $0.09\pm0.01$ & N/A & Unconstrained & 3.11\\
\hline
Case 2 ($n=0$) & $0.33\pm0.04$ & $-0.06\pm0.04$ & Unconstrained & 0.62 \\
\hline
Case 3 (Eq. \ref{nwrelation}) & $0.41^{+0.06}_{-0.16}$ & $-0.10^{+0.07}_{-0.04}$ & Unconstrained & 0.62 \\
\hline
Case 4 ($n=1$) & $0.25^{+0.15}_{-0.04}$ & $0.01^{+0.06}_{-0.20}$ & $-0.09^{+0.18}_{-0.03}$ & 0.64 \\
\hline
\end{tabular}
\end{center}
\end{table*}

\begin{figure*}
\begin{center}
\includegraphics[width=\columnwidth]{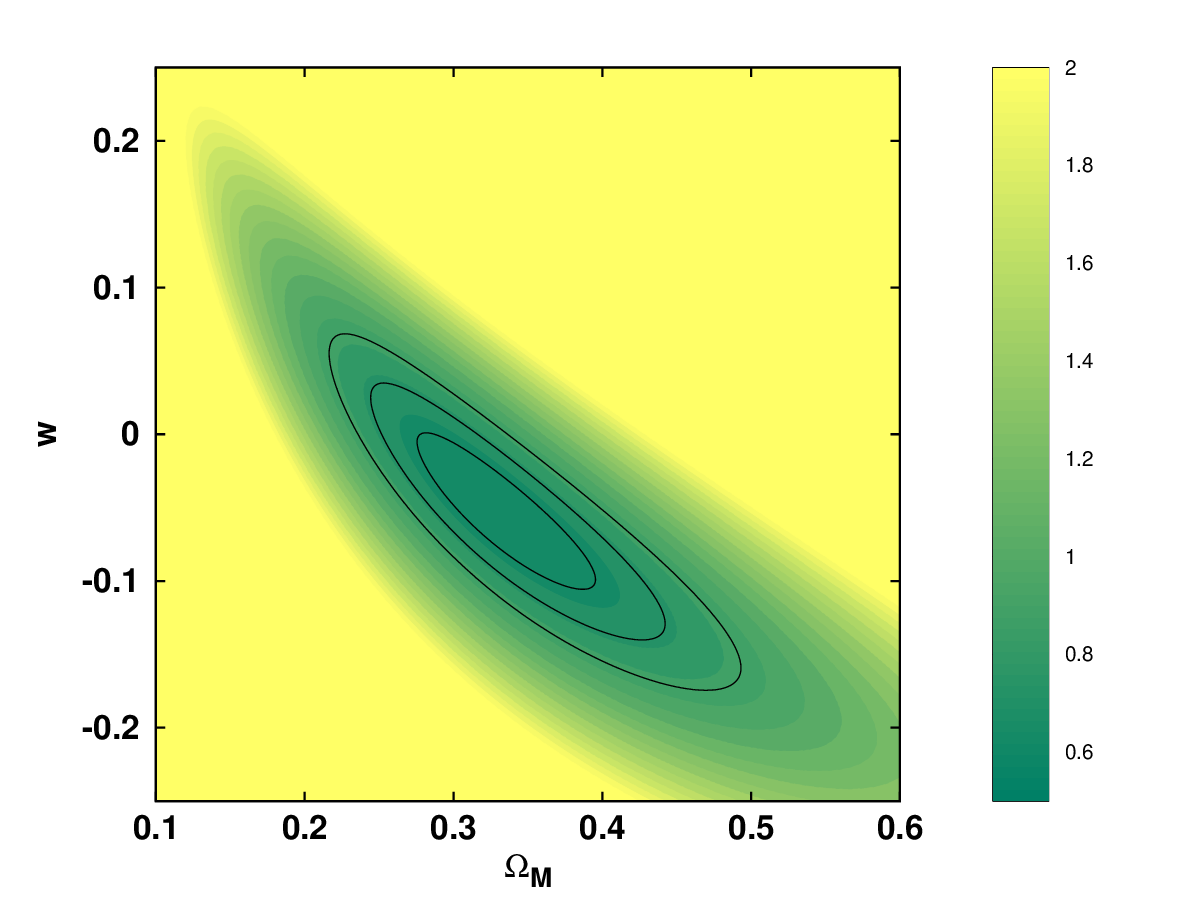}
\includegraphics[width=\columnwidth]{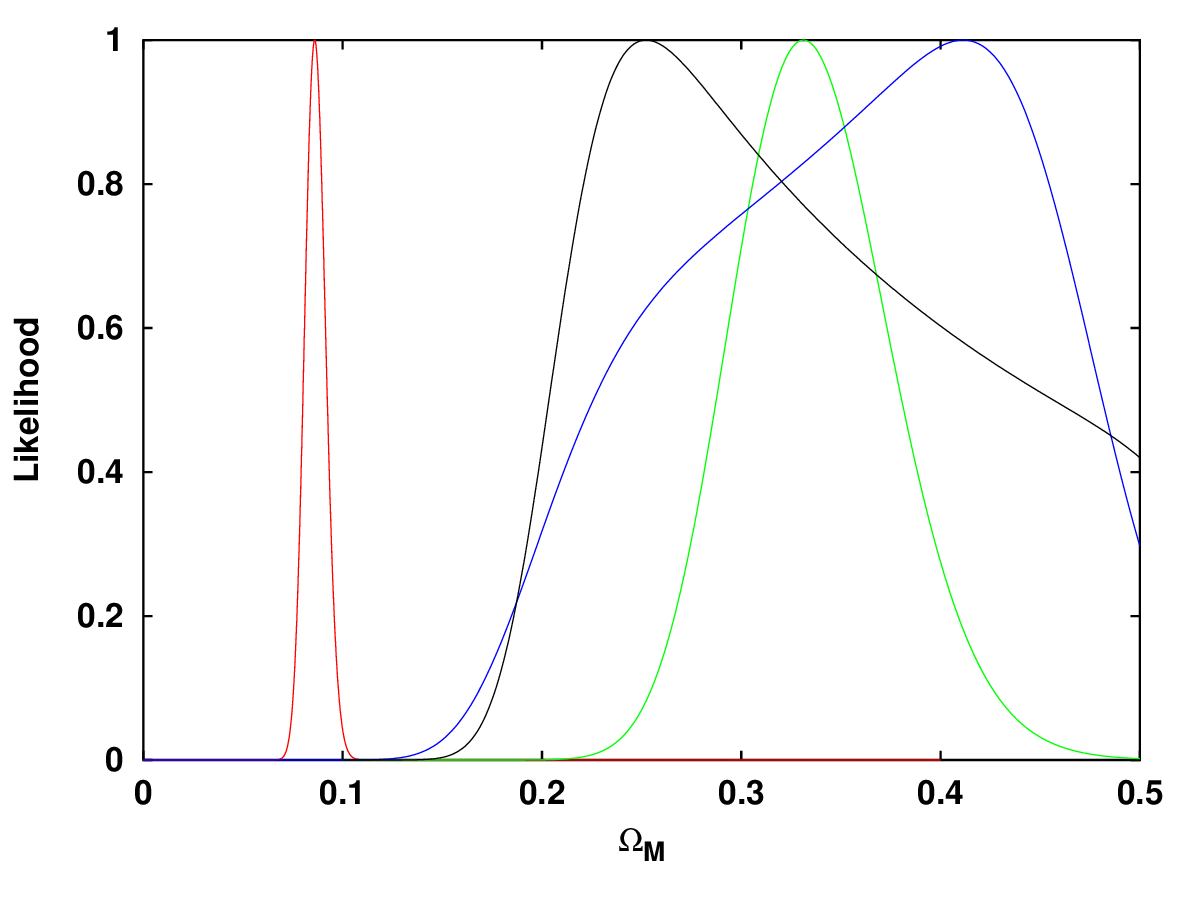}
\includegraphics[width=\columnwidth]{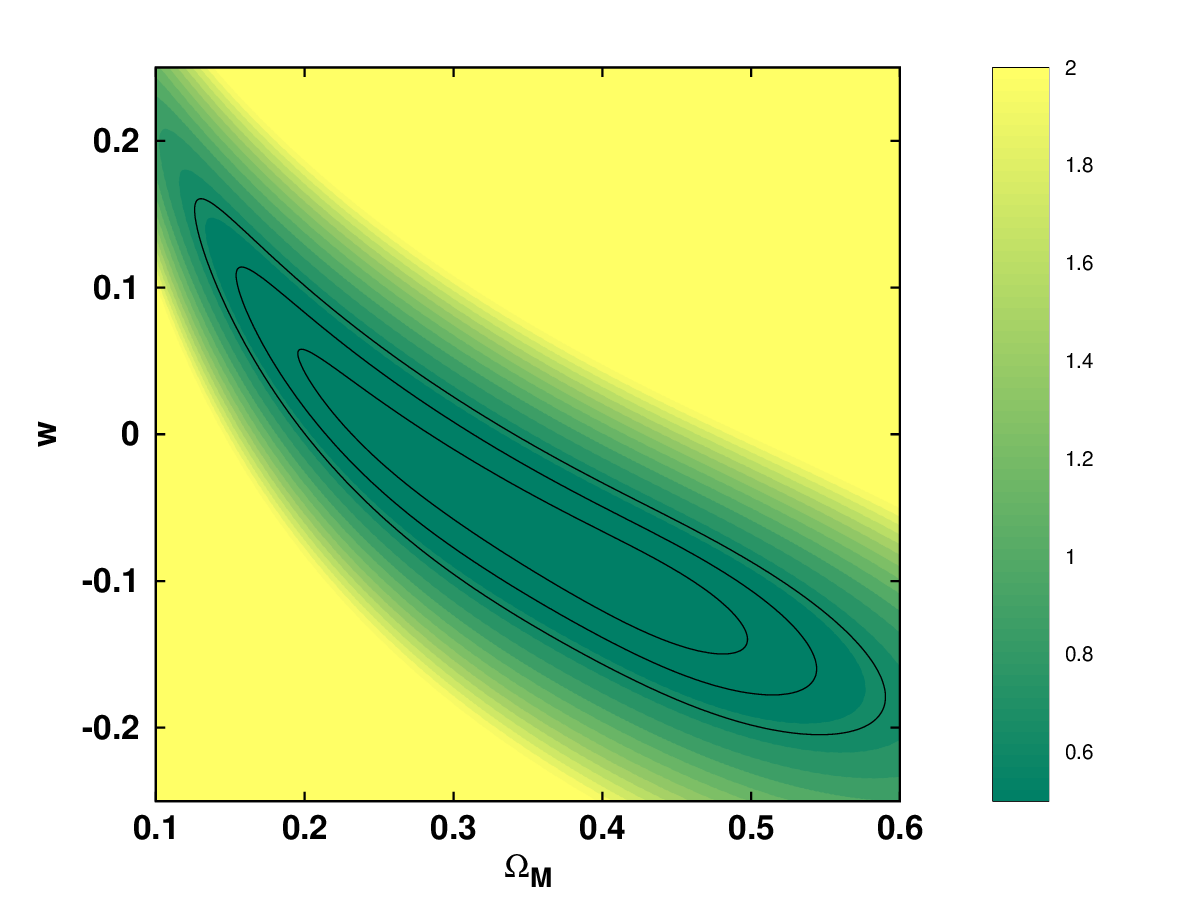}
\includegraphics[width=\columnwidth]{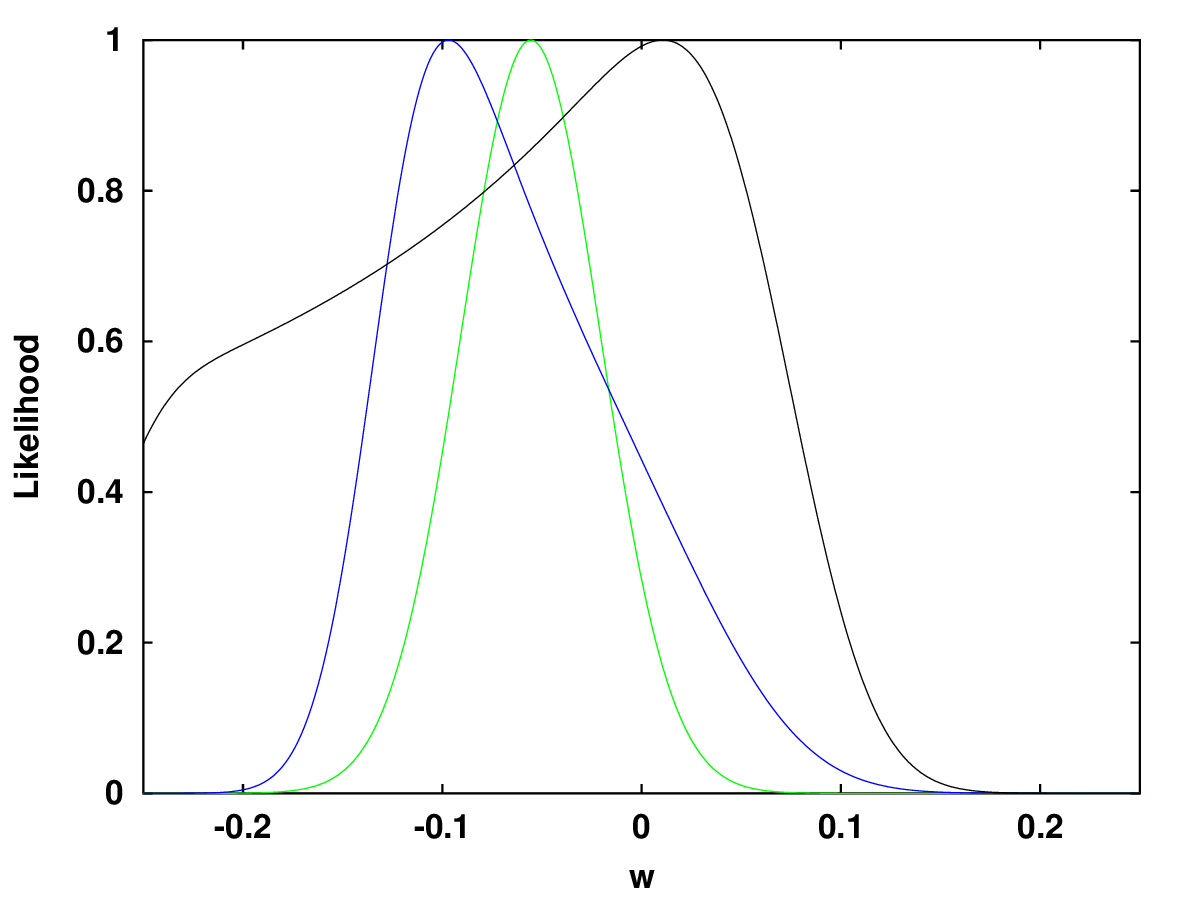}
\includegraphics[width=\columnwidth]{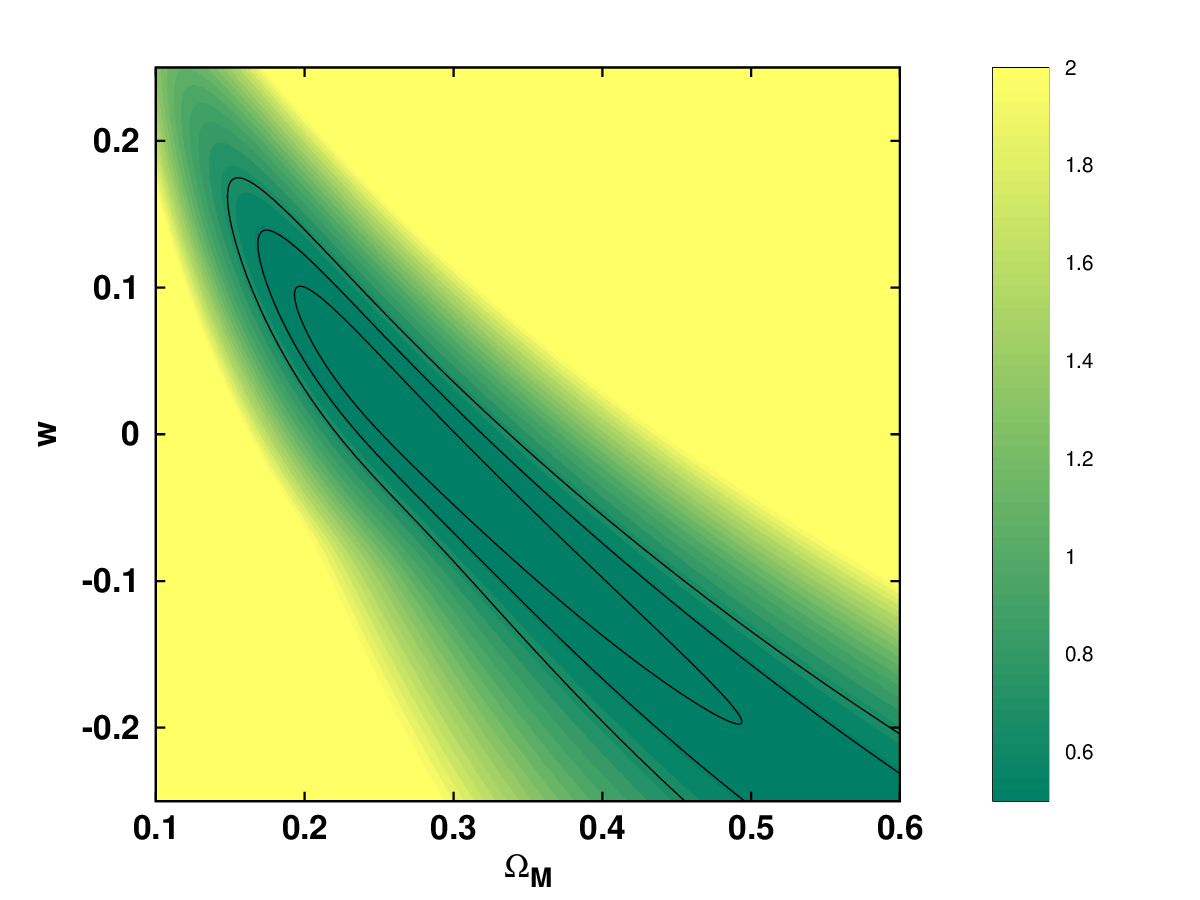}
\includegraphics[width=\columnwidth]{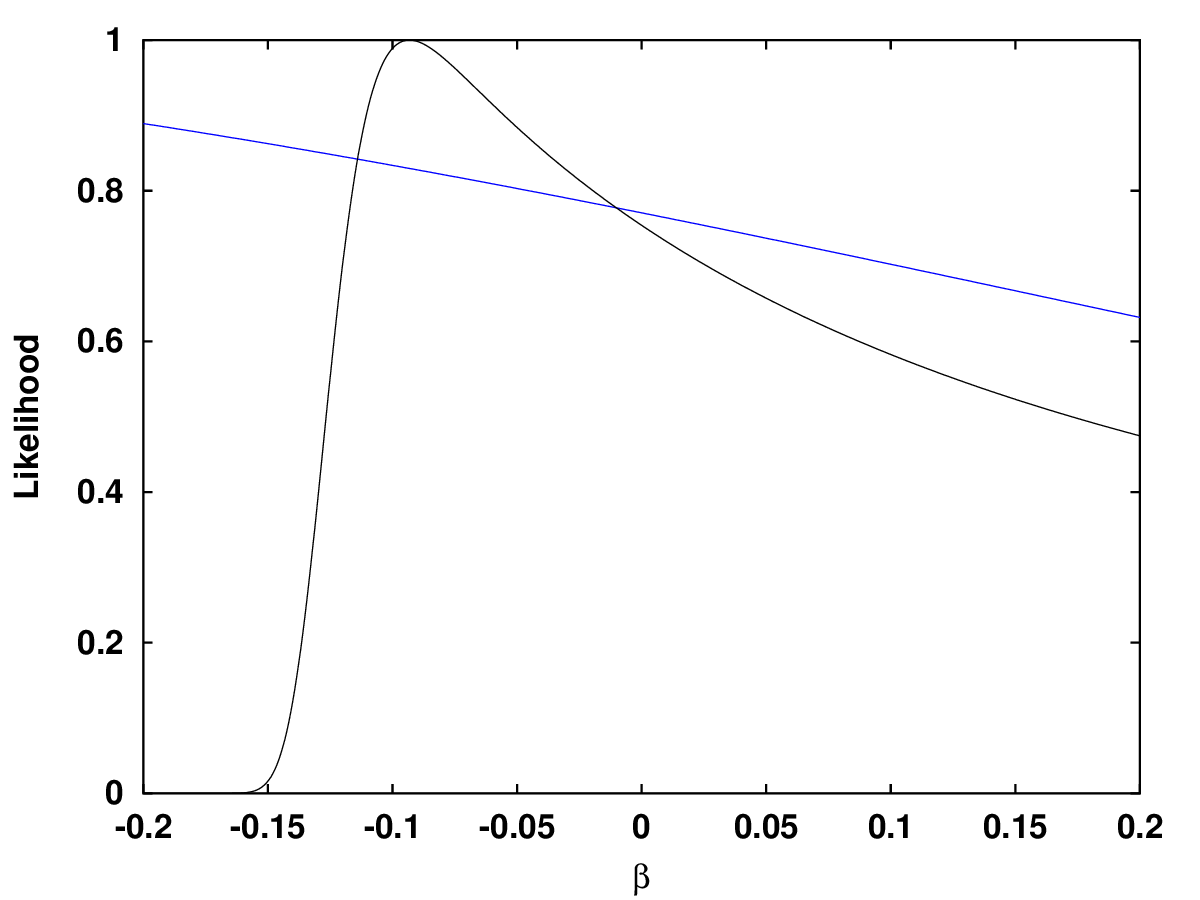}
\end{center}
\caption{Constraints on the four particular cases of the $T^n$ model discussed in Sect. \ref{partconst}. Left-side panels show constraints in the $\Omega_M$--$w$ plane for Cases 2, 3 and 4 (top, middle and bottom panels respectively). Black solid curves show the one, two, and three sigma confidence levels, and color maps depict the reduced chi-square. Right-side panels show the one-dimensional posteriors for the free parameters in each of the cases. Red, green, blue and black curves correspond respectively to Case 1 ($w=1/3$), Case 2 ($n=0$), Case 3 (Eq. \ref{nwrelation}) and Case 4 ($n=1$).}
\label{figure02}
\end{figure*}

\section{Constraints for particular cases}\label{partconst}

In this section we provide constraints on the cases, in addition to the previously discussed $\sqrt{T}$ case (in other words, the $n=1/2$ case), for which the continuity equation can be analytically integrated. Table \ref{table3} summarizes the one-sigma posterior constraints for the model parameters in each of these cases, while Fig. \ref{figure02} depicts constraints in the two-dimensional $\Omega_M$--$w$ plane, and also compares the one dimensional posterior likelihoods in the various cases, which we now introduce and discuss sequentially.

The first such case is $w=1/3$ (with any value of $n$ allowed). Here we have $r=(1+z)^{4}$ and $\Omega_\Lambda=1-\Omega_M$ so there is a single free parameter and the Friedmann equation becomes
\be
E^2(z)=(1-\Omega_M)+\Omega_M(1+z)^{4}\,.
\ee
Note that for ease of comparison with the other cases we will still denote this free parameter as $\Omega_M$, but this is obviously a radiation fluid and not a matter one. In this case we would have a universe containing radiation plus a cosmological constant, and unsurprisingly this is entirely ruled out: as listed in the first row of Table \ref{table3}, the reduced chi-square of the best-fit such model is extremely large, $\chi^2_\nu=3.11$.

The second case is $n=0$, which leads to $r=(1+z)^{3(1+w)}$, together with 
\be
\Omega_\Lambda=1-(1+\beta)\Omega_M
\ee
and therefore the Friedmann equation becomes
\be
E^2(z)=(1-\Omega_M)+\Omega_M(1+z)^{3(1+w)}\,.
\ee
Clearly this is a parametric extension of $\Lambda$CDM. The constraints are listed in the second row of Table \ref{table3}, and the corresponding $\Omega_M$--$w$ plane is shown in the top left panel of Fig. \ref{figure02}. The two parameters are anticorrelated: a matter-like fluid with a larger equation of state parameter would need to be compensated by a larger fraction of dark energy, and therefore (given our flatness assumption) by a lower matter density. In fact there is a small (one sigma, so not statistically significant) preference for a slightly larger matter density and a correspondingly negative equation of state parameter, but at the two sigma level we find compatibility with standard $\Lambda$CDM, and the model overfits the data. In any case, one should again bear in mind that there are additional constraints on the matter equation of state parameter \cite{Tutusaus}.

Perhaps more interesting is the fact that not only does $\Omega_\Lambda$ not appear explicitly in the Friedmann equation due to our flatness assumption, but the same is true for the parameter $\beta$. Phenomenologically, this means that at the background level one would have the freedom to choose the value of $\beta$ such that the cosmological constant vanishes. Specifically, and given the obtained constraint on the matter density, we have
\be
\beta_{\Lambda=0}=\frac{1-\Omega_M}{\Omega_M}=2.03\pm0.37\,.
\ee

The third case corresponds to the relation between $n$ and $w$ given by Eq. (\ref{nwrelation}), in which case the continuity equation also gives  $r=(1+z)^{3(1+w)}$. The model parameters are now related by
\be
\Omega_\Lambda=1-[1+2\beta\sigma]\Omega_M\,,
\ee
where for convenience we have defined a parameter that depends only on the equation of state parameter of the matter-like component,
\be
\sigma=(1-3w)^{(w-1)/[2(1+w)]}\,,
\ee
and the Friedmann equation has the form
\be
E^2(z)=1+\Omega_M[(1+z)^{3(1+w)}-1]+2\beta\sigma\Omega_M[(1+z)^{3(1+3w)/2}-1]\,,
\ee
This is also a two-parameter extension of $\Lambda$CDM, and therefore there are now three free parameters, $(\Omega_M, w, \beta)$. The constraints are listed in the third row of Table \ref{table3}, and the corresponding $\Omega_M$--$w$ plane is shown in the middle left panel of Fig. \ref{figure02}. In this case the parameter $\beta$ is unconstrained, while the constraints on the other two parameters are only mildly changed with respect to the previous case. In particular, the anticorrelation between these two parameters is qualitatively similar. Again, at the two sigma level we find compatibility with standard $\Lambda$CDM, and the model overfits the data.

Finally, we can also consider the case $n=1$, in which case the continuity equation does not have the standard behaviour but is still analytically integrable. Unsurprisingly, this case has some analogies to the $n=1/2$ case for the energy-momentum-powered model \cite{Faria}. Specifically, we have $r=(1+z)^{3\epsilon(1+w)}$, where for convenience we have defined 
\be
\epsilon=\frac{1+2\beta}{1+(3-w)\beta}\,,
\ee
and the model parameters are related via
\be
\Omega_\Lambda=1-[1+(3-w)\beta]\Omega_M\,.
\ee
In this case the Friedmann equation can be written
\be
E^2(z)=1+[1+(3-w)\beta]\Omega_M[(1+z)^{3\epsilon(1+w)}-1]\,,
\ee
which has the same three parameter space, $(\Omega_M, w, \beta)$, as the previous case. The constraints for this case are listed in the fourth row of Table \ref{table3}, and the corresponding $\Omega_M$--$w$ plane is shown in the bottom left panel of Fig. \ref{figure02}. Once more this is a parametric extension to $\Lambda$CDM, and the data is compatible with it. There is still an anticorrelation between the matter density and its equation of state parameter, but now the parameter $\beta$ is also constrained. In this regard, the main difference between this case and the previous one is that in the present case $\beta$ appears twice in the Friedmann equation (recall that $\epsilon$ depends on $\beta$), while in the previous case it only appeared once and was therefore strongly degenerate with the other parameters.

\begin{table*}
\begin{center}
\caption{One sigma posterior likelihoods on the matter density $\Omega_M$, power $n$, the coupling $\beta$ and the constant equation of state parameter $w$ (when applicable) for three scenarios of the full $T^n$ model, discussed in Sect. \ref{fullconst}. For the cases with $\Omega_\Lambda=0$ the constraint on $n$ is a two sigma upper limit, and the constraints on $\beta$ in brackets are not independent from those of the other parameters. The last column lists the reduced chi-square for each best-fit model. The constraints come from the combination of the Pantheon supernova data and Hubble parameter measurements.}
\label{table4}
\begin{tabular}{| c | c | c | c | c | c |}
\hline
Model assumptions & $\Omega_M$ & $n$ & $\beta$ & $w$ & $\chi^2_\nu$ \\
\hline
$\Omega_\Lambda=0$, $w=0$ & $0.22^{+0.04}_{-0.03}$ & $<0.18$ & ($3.06\pm0.07$) & N/A & $0.64$ \\
\hline
$\Omega_\Lambda\neq0$, $w=0$ & $0.26^{+0.02}_{-0.03}$ & $0.26\pm0.25$ & $0.10^{+0.35}_{-0.09}$ & N/A & $0.64$ \\
\hline
$\Omega_\Lambda=0$, $w=const.$ & $0.29^{+0.06}_{-0.10}$ & $<0.18$ & ($2.45\pm0.24$) & $-0.03^{+0.06}_{-0.05}$ & $0.64$ \\
\hline
\end{tabular}
\end{center}
\end{table*}

\section{Constraints on the full parameter space}\label{fullconst}

\begin{figure*}
\begin{center}
\includegraphics[width=0.66\columnwidth]{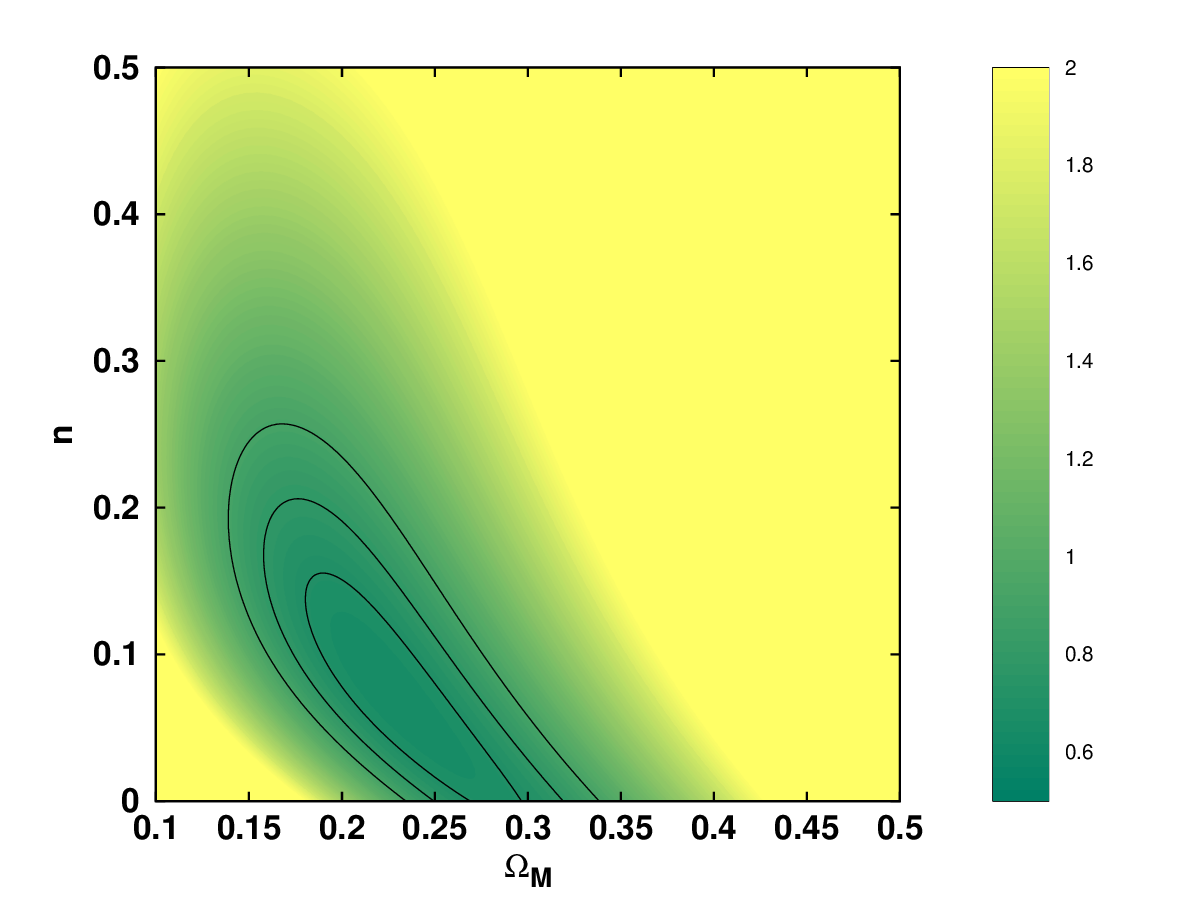}
\includegraphics[width=0.66\columnwidth]{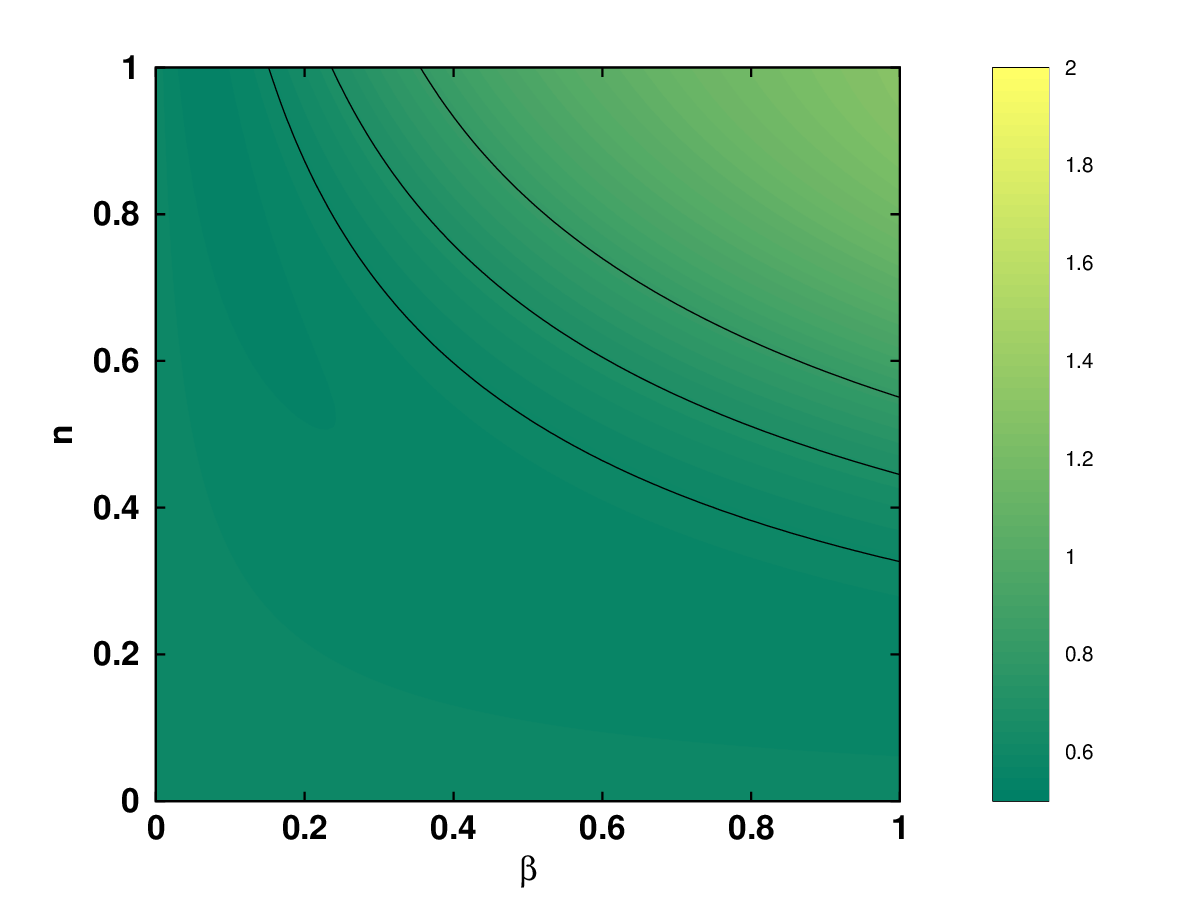}
\includegraphics[width=0.66\columnwidth]{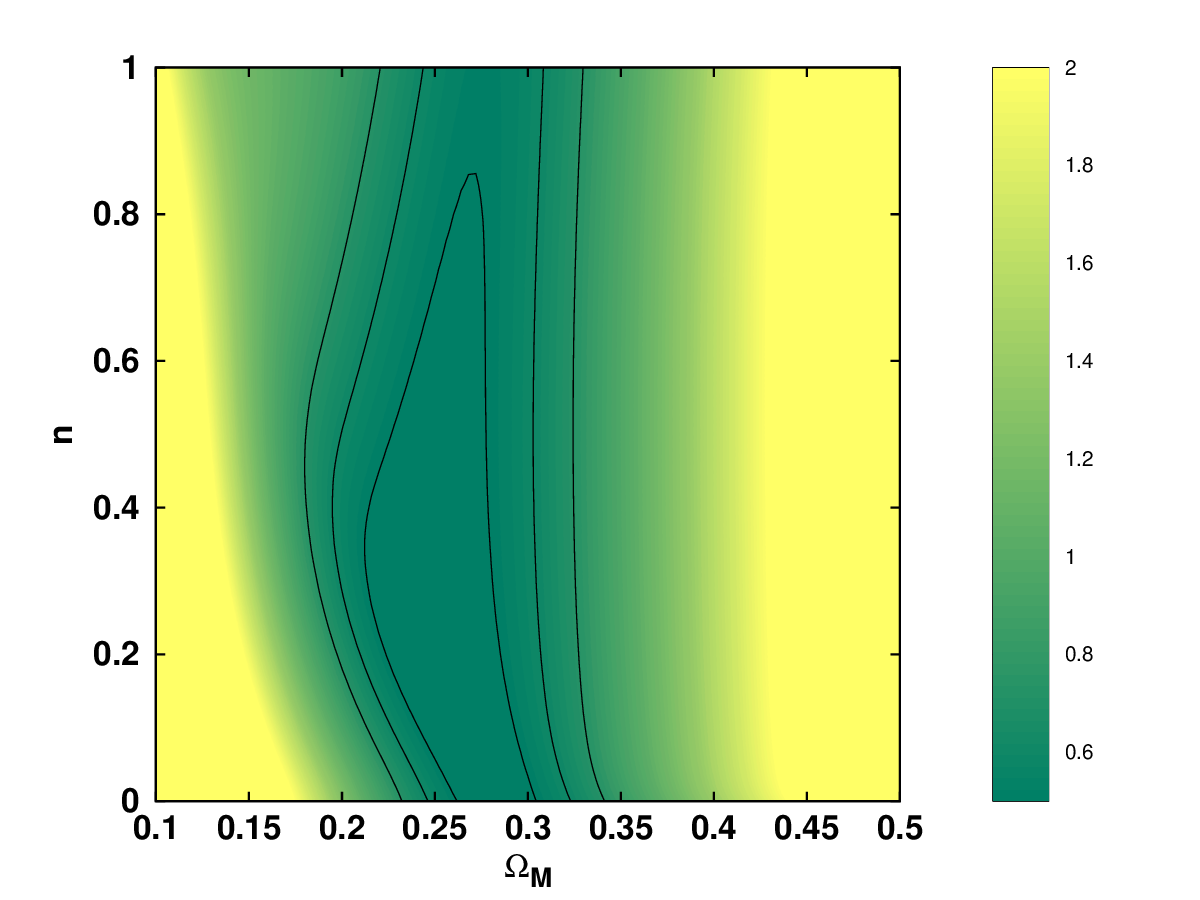}
\includegraphics[width=0.66\columnwidth]{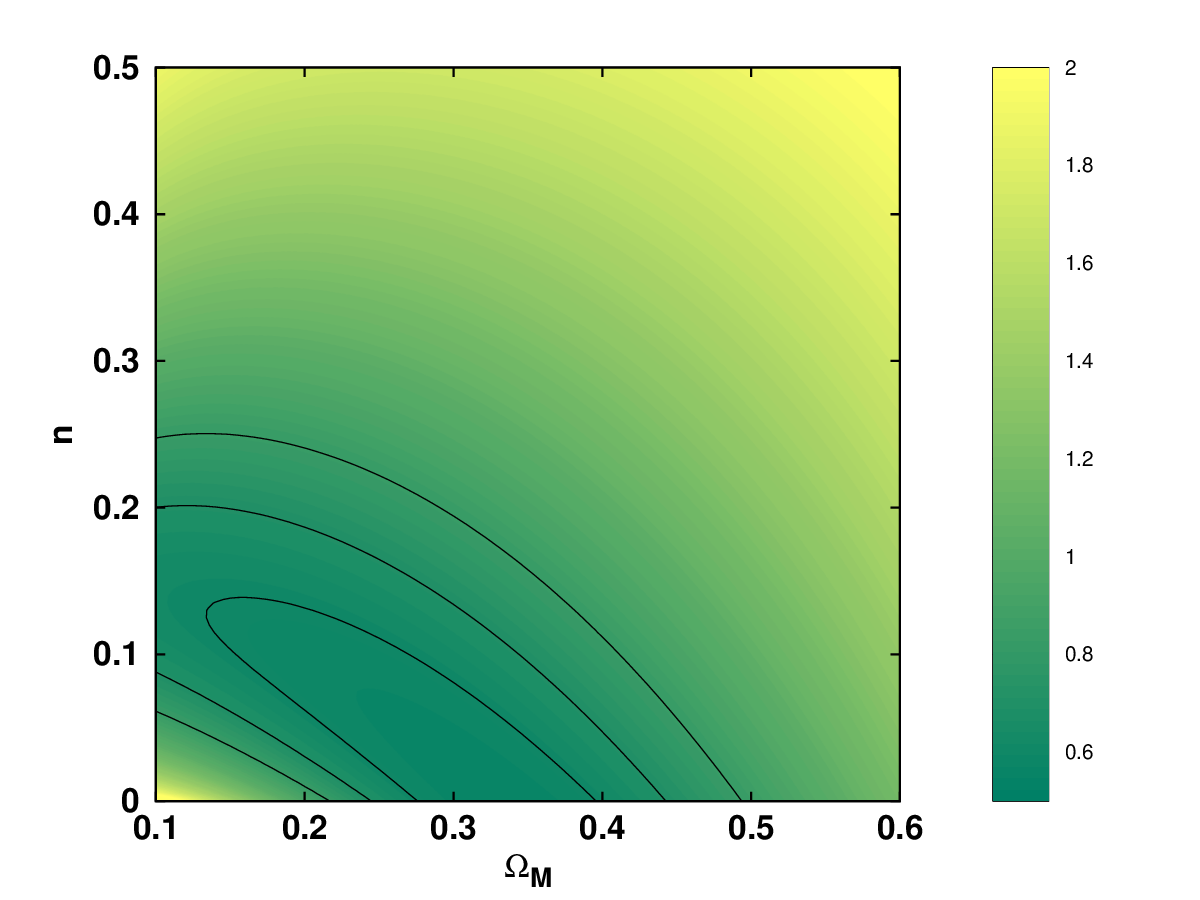}
\includegraphics[width=0.66\columnwidth]{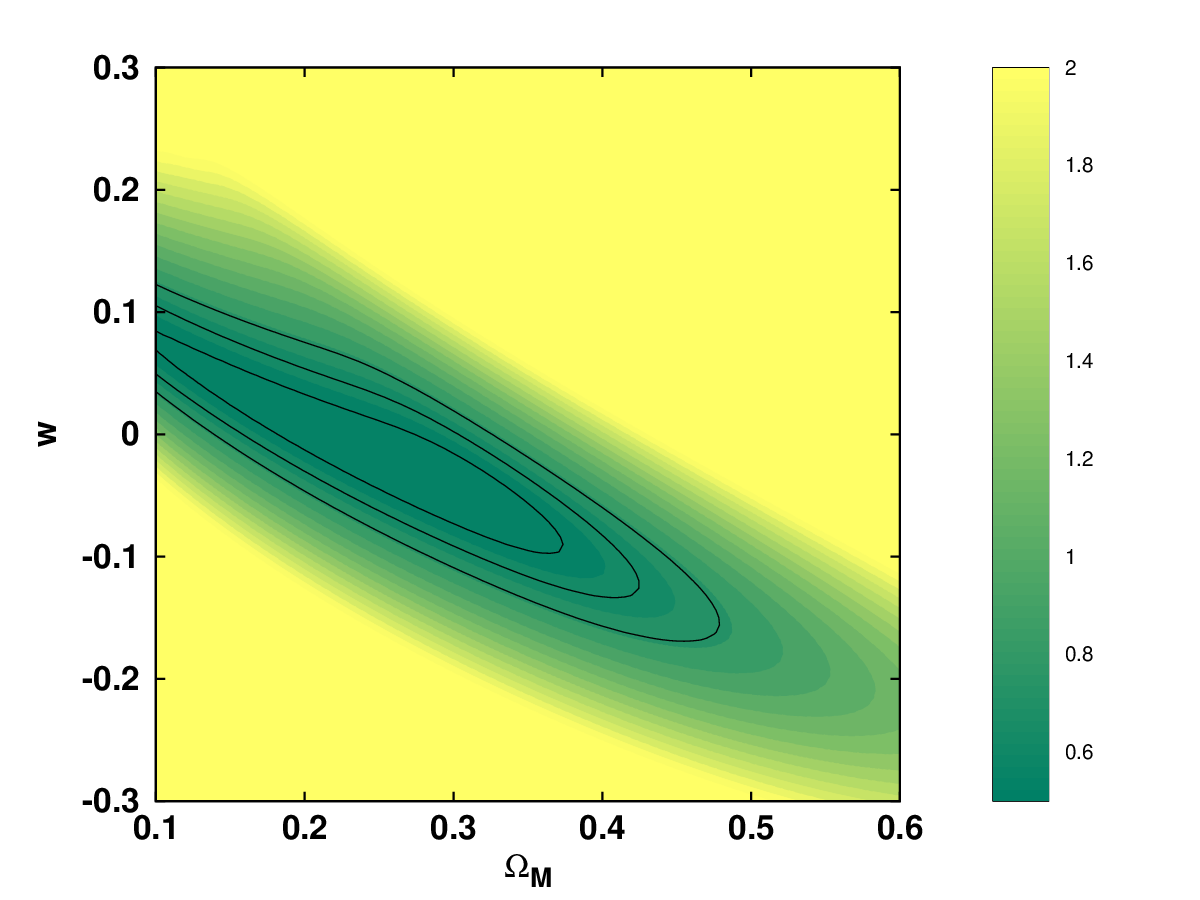}
\includegraphics[width=0.66\columnwidth]{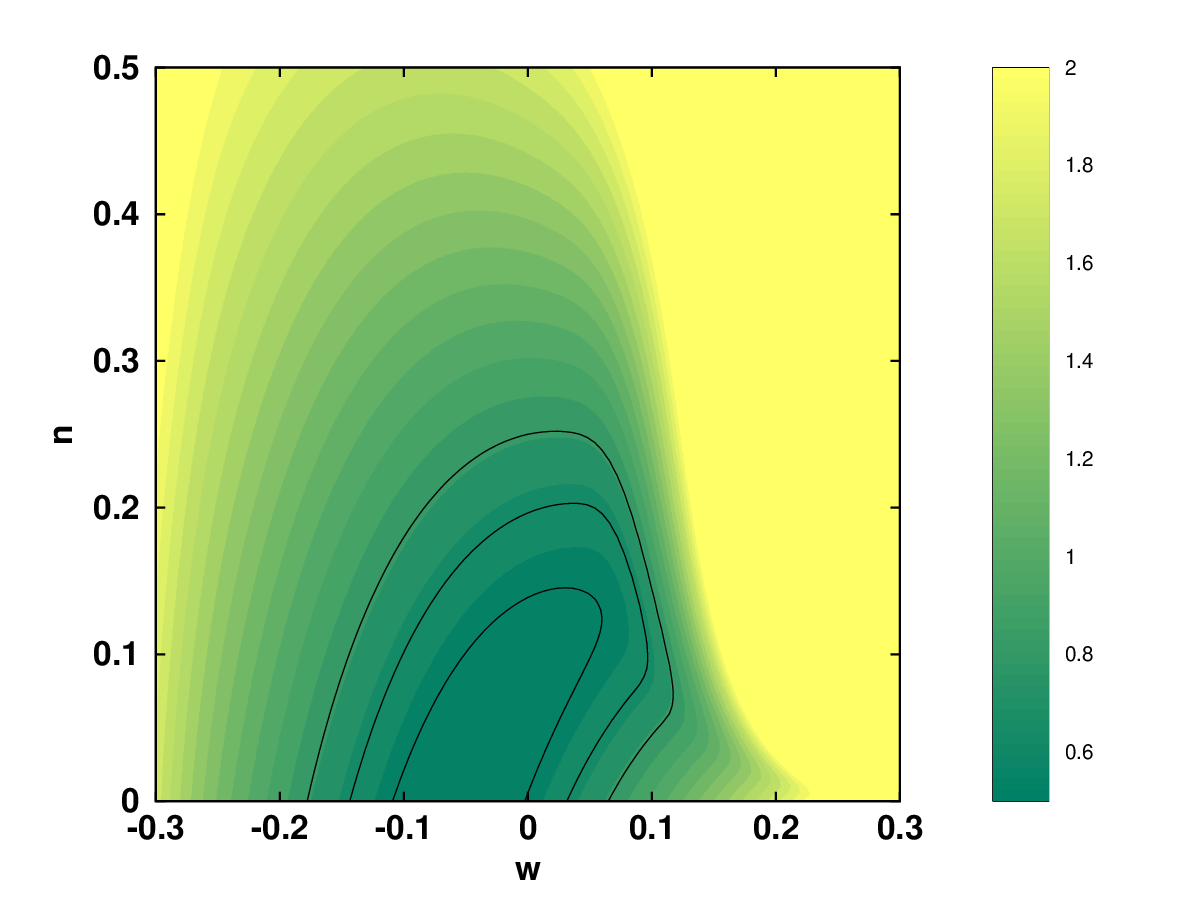}
\includegraphics[width=0.66\columnwidth]{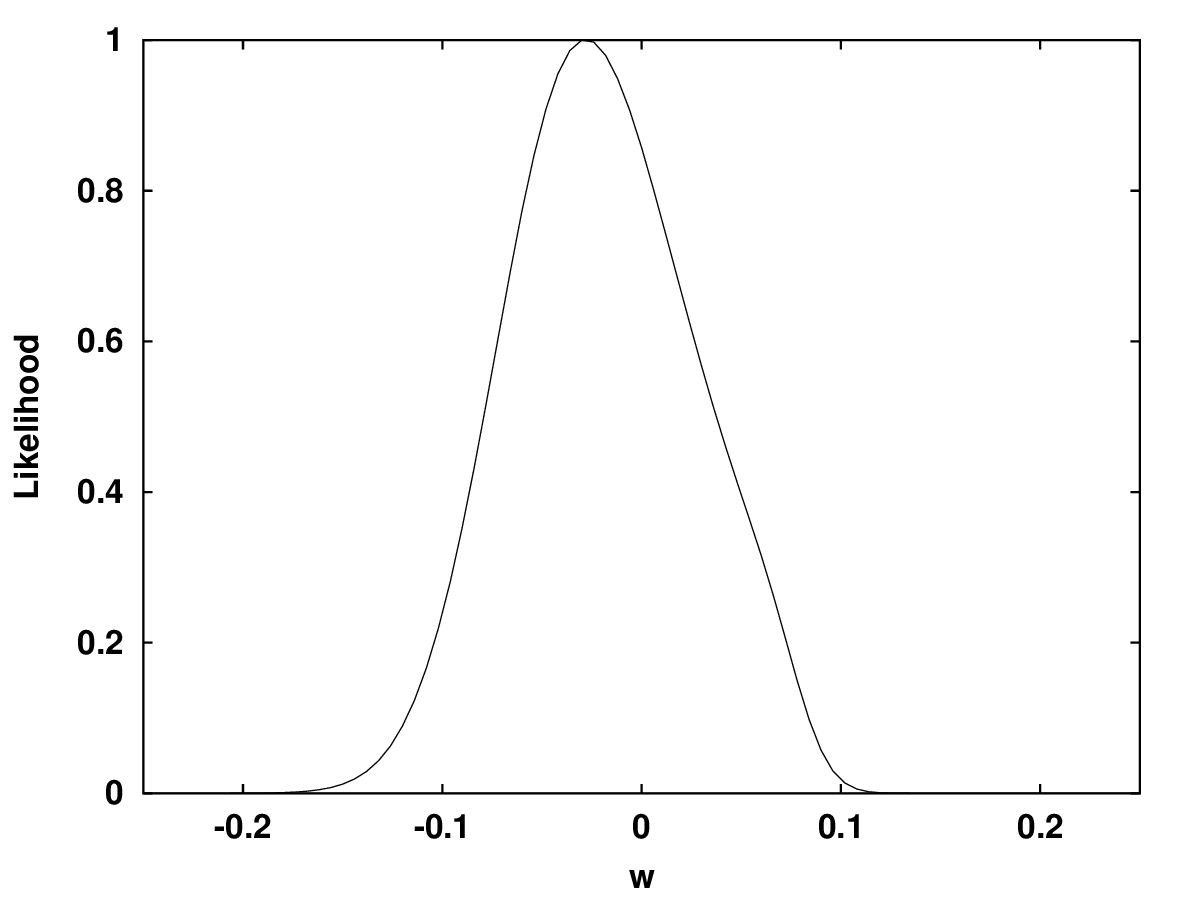}
\includegraphics[width=0.66\columnwidth]{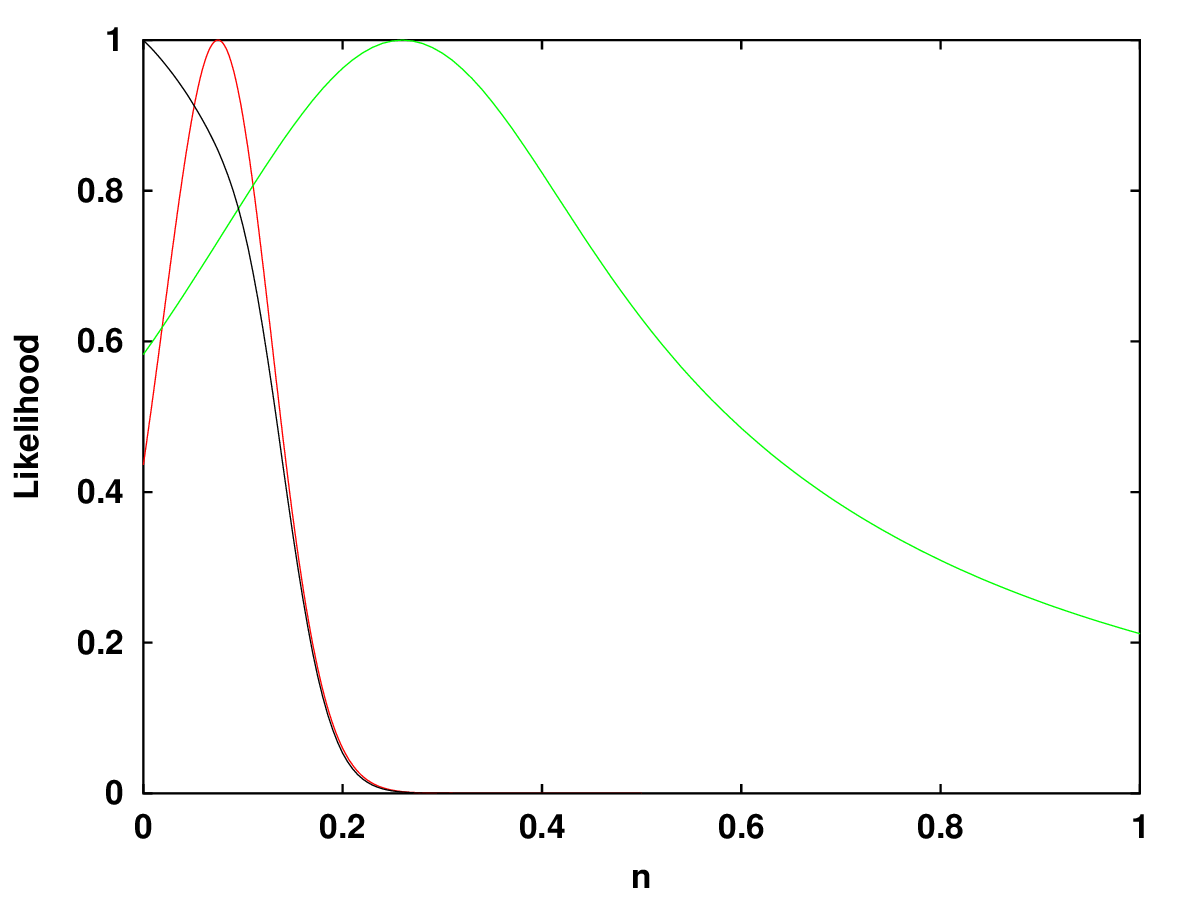}
\includegraphics[width=0.66\columnwidth]{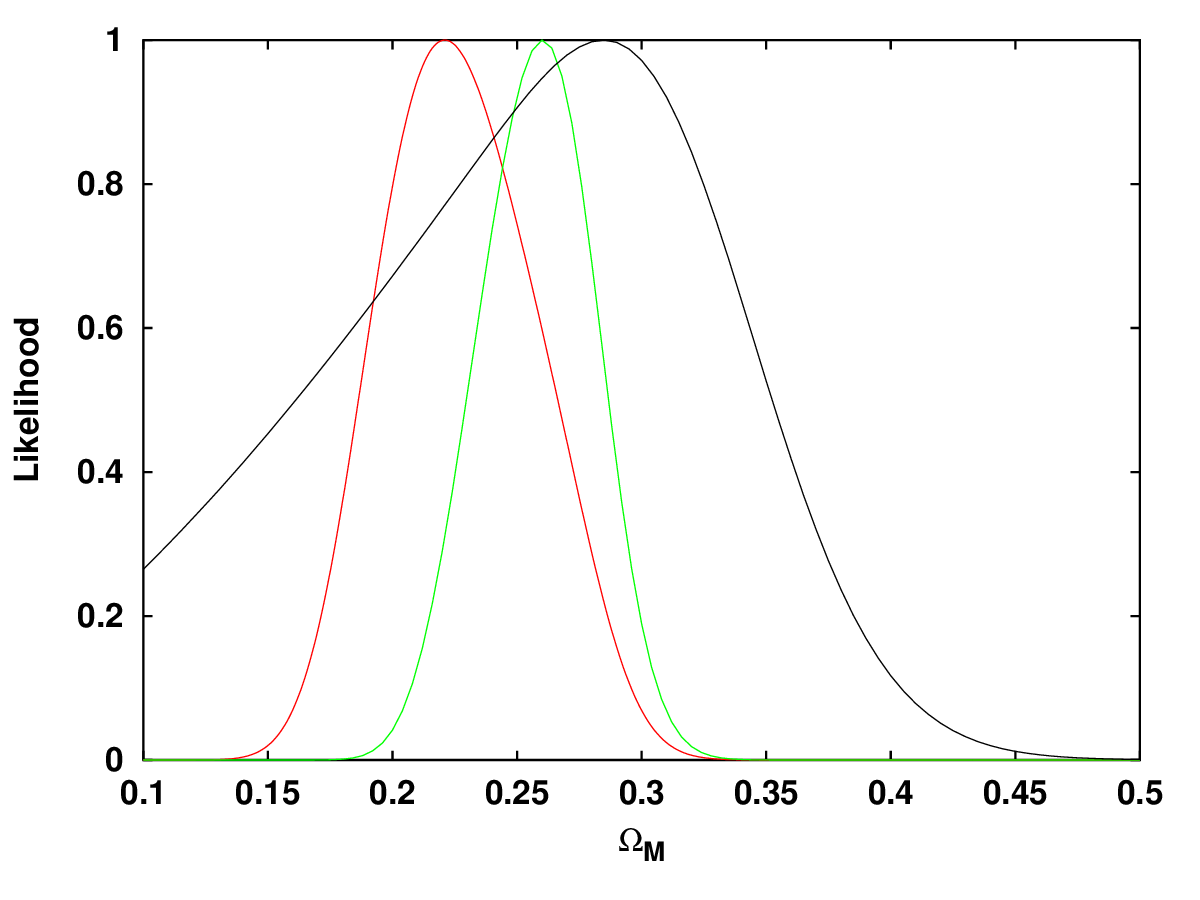}
\end{center}
\caption{Constraints on the three scenarios of the full $T^n$ model discussed in Sect. \ref{fullconst}. The top left  panel shows constraints for the ($\Omega_\Lambda=0$, $w=0)$ case, the top middle and top right panels show constraints for the ($\Omega_\Lambda\neq0$, $w=0)$ case, and the three middle row panels show constraints for the ($\Omega_\Lambda=0$, $w\neq0)$ case. In all of these the black solid curves show the one, two, and three sigma confidence levels, and the color maps depict the reduced chi-square. The bottom row panels show the one-dimensional posteriors for the free parameters (when applicable) in each of the cases. The red, green, and black curves correspond, respectively to the ($\Omega_\Lambda=0$, $w=0)$, ($\Omega_\Lambda\neq0$, $w=0)$ and ($\Omega_\Lambda=0$, $w\neq0)$ cases.}
\label{figure03}
\end{figure*}

We now discuss the constraints on the more general parameter space. In this case there is no analytic solution for the continuity equation, which must be integrated numerically. As in Sect. \ref{mods} we will do this under three separate assumptions on the presence or absence of a cosmological constant and on the allowed equation of state parameter for matter. These constraints are summarized in Table \ref{table4} and Fig. \ref{figure03}.

The first scenario has no true cosmological constant ($\Omega_\Lambda=0$), but only ordinary matter ($w=0$). Recall that we are concerned with the low-redshift behaviour of these models, and thus ignoring the radiation component. In this case there are only two free model parameters, $(\Omega_M, n)$, since the flatness condition allows $\beta$ to be expressed as
\be
\beta=\frac{1-\Omega_M}{(1+2n)\Omega_M}\,;
\ee
as expected in the absence of a cosmological constant, the $\beta=0$ case corresponds to an Einstein-de Sitter universe. Constraints on the ($\Omega_M$, $n$) plane are shown on the top left panel of Fig. \ref{figure03}. It's interesting to observe that our best fit prefers a lower than standard matter density, unlike in the analogous case for the energy-momentum powered models, cf. the first row of Table \ref{table1}. As for the constraint on $n$, one obtains a weak one sigma constraint, $n=0.08^{+0.05}_{-0.06}$. However, since a non-negative prior has been used, and bearing in mind that for $n=0$ this model reduces to $\Lambda$CDM, the constraint on $n$ is more reasonably seen as an upper limit on dynamical dark energy, at the two sigma ($95.4\%$) confidence level, one finds $n<0.18$.

The second scenario has $\Omega_\Lambda\neq0$ and $w=0$, in which case we have a three-dimensional parameter space $(\Omega_M, n, \beta)$, and constraints on two of the three two-dimensional parameter spaces (with the third parameter marginalized) are shown in the top middle and top left panels of Fig. \ref{figure03}. Clearly this is a parametric extension of $\Lambda$CDM, with two additional parameters. In this case the matter density is still tightly constrained, while $n$ and $\beta$ are very strongly degenerate and therefore much more weakly constrained. Specifically, they are constrained at one sigma but unconstrained at two sigma; in practical terms, all that one requires is that one of them is small. 

Finally, the third scenario has $\Omega_\Lambda=0$ and $w=const.$ In this case there are also three free model parameters, $(\Omega_M, n,w)$. Analogously to the first case, the flatness condition allows $\beta$ to be expressed as
\be
\beta=\frac{1-\Omega_M}{[1+2n+(2n-3)w](1-3w)^{n-1}\Omega_M}\,;
\ee
again, the $\beta=0$ case corresponds to an Einstein-de Sitter universe. The middle row panels of Fig. \ref{figure03} depict the constraints on the three relevant two-dimensional parameter spaces, with the third parameter marginalized in each case. Here our two sigma upper limit for $n$ is the same as in the first case, but the preferred value of the matter density increases with respect to that first case. This increase is not statistically significant since the corresponding error bars are also twice as large, but nevertheless the difference is mainly due to a preference for a slightly negative equation of state parameter, which again is not statistically significant.

\section{Coda: The Cardassian model}\label{Card}

In the models that we have considered in this work, the right hand side of Friedmann equation includes a nonlinear density dependent term, in addition to the standard linear one. Another example where this is assumed to happen is the so called Cardassian model of \cite{Cardassian1}, which is ostensibly an attempt to build a model of the universe that is flat, matter dominated, and accelerating, with the acceleration being due to the a nonlinear density dependent term, $\rho^n$.

Specifically, \cite{Cardassian1} assume that the low redshift universe is composed of ordinary matter (with the standard equation of state parameter, $w=0$) as well as radiation, but as usual the latter is subdominant at low redshifts. In this low redshift limit, the Friedmann equation is assumed to be
\be
E^2(z)=\Omega_M(1+z)^3+(1-\Omega_M)(1+z)^{3n}\,,
\ee
with the authors arguing that this form may be motivated by extra dimensions with time varying sizes, although this motivation has been criticized \cite{Cardassian2}. Nevertheless, we take this model at face value and constrain it using the same tools used in the rest of the article.

\begin{figure}
\begin{center}
\includegraphics[width=\columnwidth]{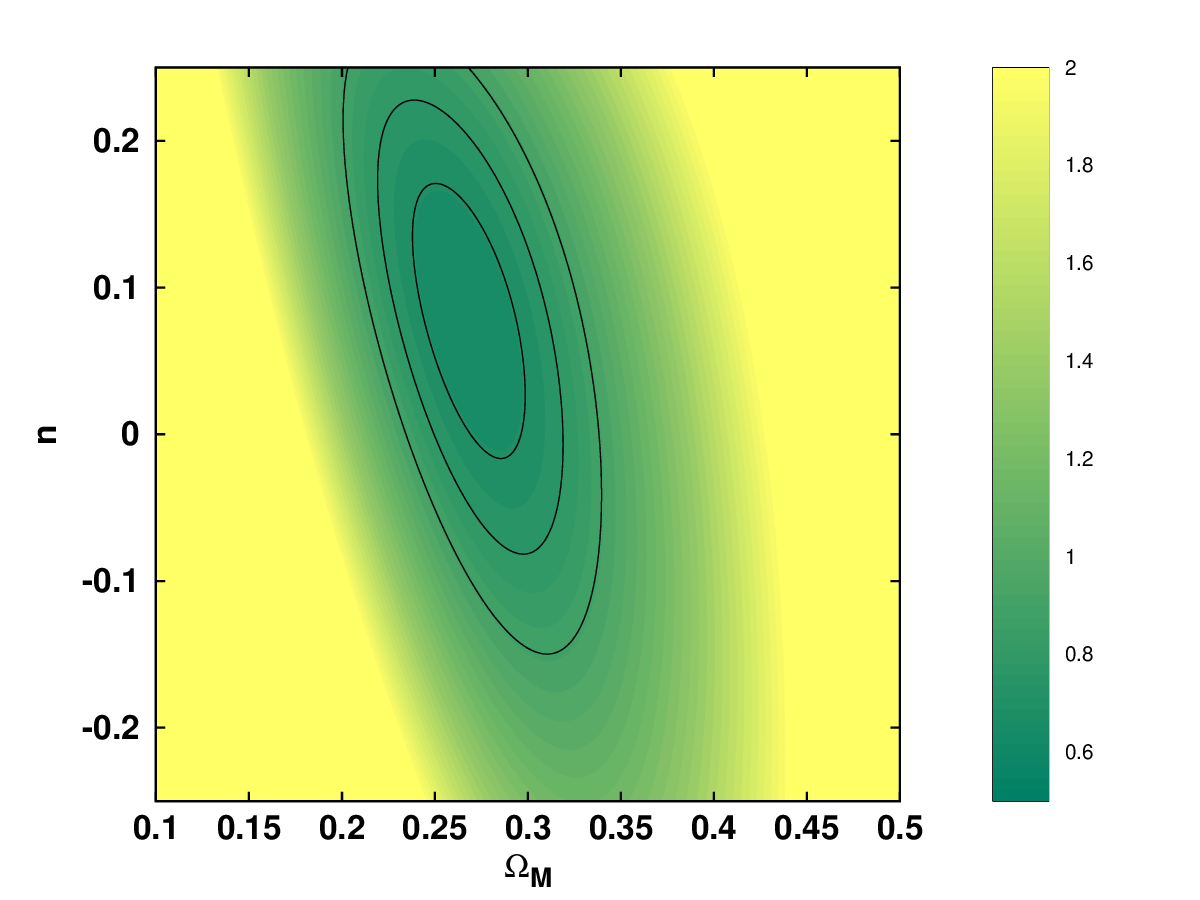}
\end{center}
\caption{Constraints on the Cardassian model, obtained from the same datasets described in the introduction and used for the other models. The black solid curves show the one, two, and three sigma confidence levels, and the color map depicts the reduced chi-square.}
\label{figure04}
\end{figure}

Constraints on the model's two-dimensional parameter space can be seen in Fig. \ref{figure04}. The one-sigma posterior likelihood constraints on the two model parameters are
\be
\Omega_M=0.27\pm0.02
\ee
\be
n=0.08\pm0.06\,;
\ee
these are consistent with $\Lambda$CDM, and also consistent with the recent analysis of \cite{Cardassian3}, when allowing for two differences between their work and ours. The first difference is that the two analyses use different supernova and Hubble parameter datasets. The second difference is that the two works make different assumptions on the Hubble constant: in our case  this parameter is always analytically marginalized, following \cite{Anagnostopoulos}, while \cite{Cardassian3} has it as an additional free parameter, for which various priors are used.

\section{Conclusions and outlook}\label{concl}

We have built upon the earlier work of \cite{Faria,Eleanna}, continuing the exploration of observational low redshift background constraints on classes of FLRW cosmological models in which the matter side of Einstein's equations includes, in addition to the canonical term, which is linearly proportional to the density, either a term proportional to a function of the energy-momentum tensor ($T^2=\rho^2+3p^2$), or to a function of its trace ($T=\rho-3p$). Both of these can be phenomenologically thought of as extensions of general relativity with a nonlinear matter Lagrangian. 

Broadly speaking, one can study the models in this class under two different scenarios. In the first one these are envisaged as phenomenological extensions of the standard $\Lambda$CDM, with one or more additional parameters. In this case the model still has a cosmological constant but the nonlinear matter Lagrangian leads to additional terms in Einstein's equations, which cosmological observations can constrain. In the second one they are considered as genuine alternatives to $\Lambda$CDM, in which there is no cosmological constant, and the nonlinear matter term would have to provide the acceleration; this scenario would be somewhat closer in spirit to the usual modified gravity models.

Overall, these three works show that parametric extensions of $\Lambda$CDM within these classes of models are tightly constrained by the datasets that we have considered, typically within one standard deviation of the canonical $\Lambda$CDM behaviour. It is also worthy of notice that these models significantly overfit the data, and in a comparative sense are poorer fits than the phenomenological CPL parametrization. This is also the case for the more ad hoc Cardassian model, briefly discussed in Sect. \ref{Card}. On the other hand, alternative models in these classes (those that do not have a $\Lambda$CDM limit) do not fit the data, and are therefore ruled out. We emphasize that our analysis only addressed low redshift background cosmology constraints. The inclusion of high redshift data, in particular from the cosmic microwave background, is left for future work, and is expected to further tighten constraints on these models. 

In conclusion, this exploration provides some insight on the level of robustness of the $\Lambda$ model and on the parameter space still available for viable alternatives and extensions. If there is no true cosmological constant, the alternative mechanism must effectively behave like one, at least at low redshifts. In the present work, this is manifest in the fact that the exponent $n$ is constrained to be close to $n=0$, in which case the nonlinear term effectively behaves as a cosmological constant. The $\Lambda$CDM paradigm is clearly a robust one. While it is manifestly a phenomenological approximation to a still unknown more fundamental model, it is also a good approximation, and any plausible alternative model must be able to closely reproduce its behaviour in a broad range of cosmological settings.

\section*{Acknowledgements}

This work was financed by FEDER---Fundo Europeu de Desenvolvimento Regional funds through the COMPETE 2020---Operational Programme for Competitiveness and Internationalisation (POCI), and by Portuguese funds through FCT - Funda\c c\~ao para a Ci\^encia e a Tecnologia in the framework of the project POCI-01-0145-FEDER-028987 and PTDC/FIS-AST/28987/2017. The project that led to this work was started during AstroCamp 2020.

\bibliographystyle{model1-num-names}
\bibliography{modgravity}
\end{document}